\documentstyle[preprint,aps,epsf]{revtex}

\def\singlespace {\smallskipamount=3.75pt plus1pt minus1pt
                  \medskipamount=7.5pt plus2pt minus2pt
                  \bigskipamount=15pt plus4pt minus4pt
                  \normalbaselineskip=12pt plus0pt minus0pt
                  \normallineskip=1pt
                  \normallineskiplimit=0pt
                  \jot=3.75pt
                  {\def\smallskip {\vskip\smallskipamount}}
                  {\def\medskip   {\vskip\medskipamount}}
                  {\def\bigskip   {\vskip\bigskipamount}}
                  {\setbox\strutbox=\hbox{\vrule
                    height10.5pt depth4.5pt width 0pt}}
                  \parskip 7.5pt
                  \normalbaselines}
\def\middlespace {\smallskipamount=5.625pt plus1.5pt minus1.5pt
                  \medskipamount=11.25pt plus3pt minus3pt
                  \bigskipamount=22.5pt plus6pt minus6pt
                  \normalbaselineskip=22.5pt plus0pt minus0pt
                  \normallineskip=1pt
                  \normallineskiplimit=0pt
                  \jot=5.625pt
                  {\def\smallskip {\vskip\smallskipamount}}
                  {\def\medskip   {\vskip\medskipamount}}
                  {\def\bigskip   {\vskip\bigskipamount}}
                  {\setbox\strutbox=\hbox{\vrule
                    height15.75pt depth6.75pt width 0pt}}
                  \parskip 11.25pt
                  \normalbaselines}
\def\doublespace {\smallskipamount=7.5pt plus2pt minus2pt
                  \medskipamount=15pt plus4pt minus4pt
                  \bigskipamount=30pt plus8pt minus8pt
                  \normalbaselineskip=30pt plus0pt minus0pt
                  \normallineskip=2pt
                  \normallineskiplimit=0pt
                  \jot=7.5pt
                  {\def\smallskip {\vskip\smallskipamount}}
                  {\def\medskip   {\vskip\medskipamount}}
                  {\def\bigskip   {\vskip\bigskipamount}}
                  {\setbox\strutbox=\hbox{\vrule
                    height21.0pt depth9.0pt width 0pt}}
                  \parskip 15.0pt
                  \normalbaselines}

\tightenlines
\begin{document}
\preprint{
\hfill$\vcenter{\hbox{\bf IUHET-424} \hbox{May
             2000}}$  }

\title{\vspace*{.75in}
Discrete Flavor Symmetries and Mass Matrix Textures}

\author{M. S. Berger
\footnote{Electronic address:
berger@gluon.physics.indiana.edu}
and Kim Siyeon
\footnote{Electronic address:
siyekim@indiana.edu}}

\address{
Physics Department, Indiana University, Bloomington, IN 47405, USA}

\maketitle

\thispagestyle{empty}

\begin{abstract}
We show how introducing discrete Abelian flavor symmetries can produce 
texture zeros in the fermion mass matrices, while preserving the correct 
relationships with the low-energy data on quark and lepton masses. We 
outline a procedure for defining texture zeros as suppressed 
entries in Yukawa matrices. These texture zeros can account for the coexistence
of the observed large mixing in atmospheric neutrino oscillations with a 
hierarchy in the neutrino masses, and offer the 
possibility of alignment of the quark and squark mass matrices, and thus 
giving a solution to the supersymmetric flavor problem. A requirement that the 
flavor symmetry commutes with the $SU(5)$ grand unified group can be used 
to explain the lepton mass hierarchies as well as the neutrino parameters, 
including the large mixing observed in the atmospheric neutrino data. 
We present one such model that yields a large atmospheric neutrino 
mixing angle, as well as a solar neutrino mixing angle of order 
$\lambda \simeq 0.22$.

\end{abstract}

\newcommand{\be}{\begin{equation}}
\newcommand{\ee}{\end{equation}}
\newcommand{\bea}{\begin{eqnarray}}
\newcommand{\eea}{\end{eqnarray}}

\newpage

\section{Introduction}
The mechanism that generates the fermion masses is not yet understood. In
the standard model (SM) the masses and mixings are simply parameters that can
be adjusted to agree with experiment. One hope is that the Yukawa couplings
in the SM can be understood more fully when the theory is embedded in a 
more fundamental theory, and relationships between masses 
and mixings might then
be established. Symmetries based on embedding the gauge symmetries of the 
SM in larger gauge groups (unified theories) have been used for a long time
and some reasonable mass patterns can be derived which are consistent with
experiment. These symmetries have come to be called vertical symmetries
to distinguish them from the horizontal symmetries (or flavor symmetries)
that relate fermions 
from the different generations. In this paper we show how discrete Abelian 
horizontal symmetries based on $Z_m$
can account for some of the successful texture patterns.

There are a number of reasons one might want to extend a $U(1)$ flavor symmetry
so that it contains an additional discrete $Z_m$ component. 
(A) The additional
discrete symmetry offers a solution to the seemingly inconsistent large mixing
observed\cite{superk} in the 
atmospheric neutrino data and a hierarchy in the  muon and tau 
neutrino masses\cite{gns}. In models with supersymmetric Abelian flavor 
symmetries, large $\nu_\mu -\nu_\tau$ mixing is achieved via a light 
neutrino mass matrix of the form\cite{gn}
\bea
&&\pmatrix{C & B\cr
                  B & A}{{v^2}\over {M}}\;,\nonumber \\
&&A,B,C\sim {\cal O}(1)\;,
\eea
where $v$ is some electroweak scale vacuum expectation value and $M$ is 
the Majorana neutrino mass scale.
The eigenvalues are typically the same order of magnitude. It requires a 
fine-tuning of the order one parameters $A$, $B$ and $C$ to achieve large
mixing between neutrinos and widely separated neutrino masses. Grossman, Nir
and Shadmi\cite{gns} advocated using a discrete symmetry to maintain the 
large mixing angle while achieving very different neutrino masses without 
fine-tuning. Discrete symmetries had been discussed previously\cite{lns}, 
but the more recent experimental results indicating large mixing in the 
atmospheric (and perhaps solar) neutrino oscillations made this technique
especially interesting. The use of a discrete flavor symmetry to understand
the mass hierarchies and mixing angles for all Standard Model fermions
was pursued in Ref.~\cite{tan,bk}. Other authors have employed non-Abelian
discrete flavor symmetries\cite{flavor,hm,fk} that have both one
and two dimensional representations. This approach is particularly well suited
for addressing the supersymmetric flavor problem where the first and second
generation of superpartners should have very similar flavor properties and 
thus belong in the same representation of the flavor group. In this paper
we restrict our attention to Abelian discrete symmetries. 
(B) The phenomenological predictions for quark mass ratios and 
Cabibbo-Kobayashi-Maskawa (CKM) matrix elements can be retained, but the 
contributions arise from a smaller number of parameters. 
(C) The process of creating the baryon asymmetry 
of the Universe by having CP-violating asymmetric decays of heavy neutrinos
can be greatly enhanced as a natural consequence of solving issue (A) 
above\cite{b-lepto}. (D) One can 
potentially solve the supersymmetric flavor problem by suppressing certain 
entries in the Yukawa matrices. The mechanism works by aligning the quark mass 
matrices with the squark mass-squared matrices\cite{ns}, and one does not 
need to require that the first and second generation squarks are degenerate. 
The mixing matrix for the squark-quark-gluino couplings can be made close to 
a unit matrix, and the undesirable flavor-changing neutral currents (FCNCs) 
are suppressed. The required suppressions are not possible with a $U(1)$ 
symmetry. 

During the last few years, there has been great interest in using new 
continuous Abelian as well as 
discrete Abelian and non-Abelian symmetries in the 
minimal supersymmetric standard model (MSSM) to describe the experimental
(phenomenological) data on the fermion masses and 
mixings\cite{gns},\cite{fn}-\cite{af}. 
Superstring theories appear to have $U(1)$ symmetries and symmetries involving
its discrete subgroups as a generic feature. 
If the $U(1)$ flavor symmetry is gauged then a general assignment of 
flavor charges to the fields will be anomalous. One can imagine the anomaly is 
canceled via the Green-Schwarz mechanism\cite{gs}, and one must check 
whether the correct relations are satisfied. A convenient 
way to ensure that the flavor charges are amenable to cancellation is to have
the flavor symmetry commute with the $SU(5)$ grand unified theory\cite{nw}. 
We present in this paper a model with a $U(1)\times Z_2$ flavor 
symmetry with at least four texture zeros (in the up and down quark 
Yukawa matrices) that commutes with the $SU(5)$ gauge group.

The paper is organized as follows. In Section II we briefly discuss flavor 
symmetries and how they can in principle account for the experimentally 
observed hierarchies in the quark masses and mixing angles. In Section III 
we discuss the possible role of a discrete component in the flavor symmetry.
Section IV then 
lists the phenomenological requirements that must be met in the quark sector
of the standard model. 
In Section V a particular model for which the flavor symmetry commutes with
an $SU(5)$ grand unified symmetry is presented, and the phenomenology is 
extended to include the leptons. The consequences for neutrino oscillations 
and the charged lepton masses are discussed. Finally we present our conclusions
in Section VI. 

\section{Flavor Symmetries}
The hierarchical structure of the fermion mass matrices hints
that there may be a spontaneously broken family symmetry responsible for the 
suppression of Yukawa couplings.
In this paper we employ supersymmetric Abelian horizontal symmetries.
These flavor symmetries allow the 
fermion mass and mixing hierarchies to be naturally 
generated from nonrenormalizable 
terms in the effective low-energy theory.

The idea is quite simple and easily implemented\cite{fn}. 
There is some field $S$
which is charged under a $U(1)$ family symmetry, and without loss of 
generality, we can assume that its charge is -1. There are terms contributing 
to effective Yukawa couplings for the quarks,
\bea
&&Q_i\overline{d}_jH_d\left ({{S}\over {\Lambda _L}}\right )^{m_{ij}}
+Q_i\overline{u}_jH_u\left ({{S}\over {\Lambda _L}}\right )^{n_{ij}}\;,
\eea
and the integer exponents $m_{ij}$ and $n_{ij}$ 
are easily calculated in terms of the 
horizontal symmetry charges of the quark and Higgs fields. 
The scale, $\Lambda _L$, where massive states are integrated out of the 
fundamental theory to produce an effective theory, is assumed to be 
larger than the vev, $<S>$ of the singlet scalar field so the parameter
$<S>/\Lambda _L$ is a small one.  We henceforth require the
Higgs fields to be uncharged under the $U(1)$ family symmetry, 
then the exponent 
$m_{ij}$ is just the sum of the horizontal charge of the fields 
$Q_i$ and $\overline{d}_j$. 
The hierarchy is generated from terms 
in the superpotential that carry integer 
charges $m_{ij},n_{ij}\geq 0$. If we call the small breaking parameter 
$<S>/\Lambda _L\sim \lambda$, then the generated terms for say the 
down quark 
Yukawa matrix will be of order $\lambda ^{m_{ij}}$.
We will restrict our attention in this 
paper to flavor charges for the Standard Model fields that are non-negative.
Here texture zeros refer to Yukawa matrix 
elements that can be replaced by an exact zero without 
affecting the leading order (in the small parameter $\lambda$) results for 
the mass eigenvalues and mixing angles.
An analysis of the possible approaches to explaining
the neutrino masses and mixings using $U(1)$ symmetries only is given in 
Ref.~\cite{lr}.

In models whose flavor symmetry contains two distinct components 
($U(1)\times U(1)$, $U(1)\times Z_m$, etc.) one introduces\cite{lns} 
two singlet scalars,
$S_1$ and $S_2$, with horizontal charges
\bea
&&S_1(-1,0)\;, \qquad S_2(0,-1)\;,
\eea
which in general can have different vacuum expectation values $<S_1>$ and 
$<S_2>$. These can be related to a common expansion parameter $\lambda$ by 
setting 
\bea
{{<S_1>}\over {\Lambda _L}}\sim \lambda ^\beta\;, \qquad
{{<S_2>}\over {\Lambda _L}}\sim \lambda ^\alpha\;.
\eea
In the following we identify $\lambda $ as the Cabibbo angle, and take 
$\beta =1$. In general, one can take $\alpha \ne 1$, but for our explicit 
models we assume $\alpha =1$. The contributions to the Yukawa matrices arise
from flavor invariant terms in 
\bea
&&Q_i\overline{d}_jH_d\left ({{S_1}\over {\Lambda _L}}\right )^{m_{ij}}
\left ({{S_2}\over {\Lambda _L}}\right )^{p_{ij}}
+Q_i\overline{u}_jH_u\left ({{S_1}\over {\Lambda _L}}\right )^{n_{ij}}
\left ({{S_2}\over {\Lambda _L}}\right )^{q_{ij}}\;. \label{fnmech}
\eea
It should be understood that there are undetermined order one coefficients
multiplying these terms, and we assume in this paper that these coefficients
are sufficiently close to one so as not to influence the hierarchy, i.e.
somewhat greater than $\lambda $ and somewhat less than $1/\lambda$.
Formulae for the Yukawa matrices for the quarks and charged leptons as well
as mass matrices for the neutrinos that follow from the Froggatt-Nielsen 
mechanism are given in the Appendix.

If the flavor symmetry is $U(1)$ then there is a charge assignment in
the quark sector that satisfies all the phenomenological 
requirements detailed in Section IV below. This solution was 
obtained by many authors\cite{bd,eir,irges}. The up and down quark Yukawa
matrices are
\bea
&&{\bf U}\sim \pmatrix{\lambda ^8 & \lambda ^5 & \lambda ^3 \cr
                 \lambda ^7    & \lambda ^4 & \lambda ^2 \cr
                 \lambda ^5 & \lambda ^2 & 1}\;,\qquad
  {\bf D}\sim \pmatrix{\lambda ^4 & \lambda ^3 & \lambda ^3\cr
                 \lambda ^3 & \lambda ^2 & \lambda ^2\cr
                    \lambda & 1 & 1}\;. \label{eir}
\eea 
The model we present in this paper will give the same phenomenological 
predictions as Eq.~(\ref{eir}), 
but the discrete symmetry will suppress
certain entries in comparison to the $U(1)$ flavor symmetry pattern shown in
Eq.~(\ref{eir}). After including a discrete component to the flavor symmetry, 
a different $SU(5)$ grand unified model can be constructed (see Section V).

\section{Discrete Abelian Groups}

In this section we discuss the two possible texture patterns for a $2\times 2$ 
matrix, and then show how to put these $2\times 2$ blocks together to 
form the realistic case of texture patterns for three generations.

\subsection{Suppressing in $2\times 2$ blocks}

When the flavor symmetry is $U(1)$, there is a sum rule among the exponents
in any $2\times 2$ block. For example, the up quark Yukawa matrix 
necessarily has
the relationship, 
\bea 
n_{ii}+n_{jj}-n_{ij}-n_{ji}=0\; \label{rule}
\eea
between the exponents, $n_{ij}\equiv q_i+u_j$. The Yukawa matrices in 
Eq.~(\ref{eir}) obey this rule, for example.
However, these relationships between elements of the Yukawa matrices 
can be avoided if the flavor symmetry has a $Z_m$ component. 
We can illustrate this with a simple example with a $Z_2$ symmetry:
Consider two generations with $Z_2$ flavor charges as follows,
\bea
&& Q_L: \quad q^Z_i=(0,1)\;, \qquad \overline{u}_R: \quad u^Z_i=(0,1)\;, 
\qquad i=1,2\;, \label{particular}
\eea
where the first number for each field gives the charge for the first generation
and the second number gives the charge for the second generation.
Then performing the $Z_2$ arithmetic in constructing the contribution to the 
Yukawa matrices yields, in general, 
\bea
&&\pmatrix{\lambda ^{[q^Z_1+u^Z_1]} & \lambda ^{[q^Z_1+u^Z_2]} \cr
                  \lambda ^{[q^Z_2+u^Z_1]} & \lambda ^{[q^Z_2+u^Z_2]}}\;.
\label{u23}
\eea
We use brackets around the exponents, 
$[\:\:\: ]$, to denote that we are modding out by two 
according to the $Z_2$ addition rules.
In the case of the particular choice of charges in Eq.~(\ref{particular})
and taking $<S_2>/\Lambda _L\sim \lambda $
\bea
&&\pmatrix{ \lambda ^0 & \lambda ^1 \cr
            \lambda ^1 & \lambda ^0 }\;,
\label{u23new}
\eea
So this set of charges yields a Yukawa matrix that does not satisfy the rule 
in Eq.~(\ref{rule}). If one adds in nontrivial contributions from the $U(1)$ 
part of the flavor symmetry, one sees that the off-diagonal entries in 
Eq.~(\ref{u23new}) are suppressed relative to the expectation from 
Eq.~(\ref{rule}). For example, assume that the fields have (in addition to the 
$Z_2$ assignments in Eq.~(\ref{u23new})) the $U(1)$ charge assignments
\bea
&& Q_L: \quad q_i=(3,0)\;, \qquad \overline{u}_R: \quad u_i=(1,0)\;, 
\qquad i=1,2\;, \label{particular2}
\eea
which, in general, give a contribution to the Yukawa matrices 
\bea
&&\pmatrix{\lambda ^{q_1+u_1} & \lambda ^{q_1+u_2} \cr
                  \lambda ^{q_2+u_1} & \lambda ^{q_2+u_2}}\;.
\label{u23x}
\eea
The particular choice of charges in Eq.~(\ref{particular2})
together with 
taking $<S_1>/\Lambda _L\sim \lambda $ yields the contribution to 
the Yukawa matrices from the $U(1)$ charges of the form
\bea
&&\pmatrix{ \lambda ^4 & \lambda ^3 \cr
            \lambda ^1 & \lambda ^0 }\;.
\label{u23new2}
\eea
It should be clear from Eq.~(\ref{fnmech}) that the overall contribution
to the Yukawa matrix is the product of the contribution from the $Z_2$ charges
in Eq.~(\ref{u23new}) and the contribution from the $U(1)$ charges in 
Eq.~(\ref{u23new2}) for each element of the Yukawa matrix. For the example here
we get the following result for the $U(1)\times Z_2$ flavor symmetry,
\bea
&&{\bf U}\sim \pmatrix{ \lambda ^4 & \lambda ^4 \cr
            \lambda ^2 & \lambda ^0 }\;.
\label{u23new3}
\eea
So one sees that the off-diagonal entries are suppressed relative to the 
expectation in Eq.~(\ref{u23new2}), and it is not difficult to convince oneself
that this suppression comes entirely from the $Z_2$ part of the flavor 
symmetry.
 
The relevance of the above example to the present paper is the following. We
are interested in determining the phenomenological predictions of the Yukawa
matrices and then comparing them to the experimental data. This requires that
we diagonalize the Yukawa matrices to determine the eigenvalues (masses) and 
the mixing angles (CKM elements) as explained in the Appendix. 
In the example we arrived at the 
Yukawa matrix in Eq.~(\ref{u23new3}), for which it is immediately clear 
that the eigenvalues are of order $\lambda ^0$ and $\lambda ^4$, while the 
mixing angles in $V_u^L$ and $V_u^R$ (see Eq.~(\ref{biunit})) are of order 
$\lambda ^4$ and $\lambda ^2$, respectively. 
So one notes that the left-handed mixing angle is suppressed by 
$\lambda^2$ in comparison to the expectation from a $U(1)$ symmetry 
alone that gives the same mass eigenvalues, namely
\bea
&&\pmatrix{ \lambda ^4 & \lambda ^2 \cr
            \lambda ^2 & \lambda ^0 }\;.
\label{u1only}
\eea
One can compare this simple
$2\times 2$ example with the second and third generations of Eq.~(\ref{eir}). 
We denote this suppression in the following way,
\bea
\pmatrix{X & 0 \cr 0 & X}\;.
\eea
We say that the Yukawa matrix has texture zeros in the off-diagonal positions.
A texture zero defined in this way is not a true zero, but is negligible 
to the leading order in the small parameter $\lambda $ as far as the mass
eigenvalues and the left-handed mixing (diagonalization) angles are concerned.
The right-handed mixing angle is {\it not} suppressed but this affects 
neither the CKM mixing angles nor the mass eigenvalues.
The physical observables are the elements of the CKM matrix, 
Eq.~(\ref{ckmdef}), and involve contributions from the diagonalization of 
the down-Yukawa matrix as well. So if there is a contribution from the 
diagonalization of down-Yukawa matrix of order $\lambda $, the contribution 
from the up-Yukawa matrix will be negligible in comparison. In this case
we promote the texture zero to a true phenomenological zero: the contribution
from the off-diagonal elements of the up-quark Yukawa matrix does 
not contribute
to the determination of any physical observable (quark mass or CKM element) 
to leading order in the expansion in the small parameter $\lambda $.

We can also engineer a suppression along the diagonal elements of a Yukawa 
matrix. This case is somewhat trickier than the previous case, so we proceed 
now to present another example in the case of just two generations:
For the $Z_2$ charges, consider the assignment
\bea
&& Q_L: \quad q_i^Z=(1,0)\;, \qquad \overline{d}_R: \quad d_i^Z=(0,1)\;, 
\qquad i=1,2\;. \label{particular3}
\eea
and, for the $U(1)$ charges, make the assignment
\bea
&& Q_L: \quad q_i=(3,0)\;, \qquad \overline{d}_R: \quad d_i=(1,0)\;, 
\qquad i=1,2\;. \label{particular4}
\eea
Then one obtains 
\bea
&&\pmatrix{ \lambda ^1 & \lambda ^0 \cr
            \lambda ^0 & \lambda ^1 }\;.
\label{d23new4}
\eea
for the contribution from the $Z_2$ charges and
\bea
&&\pmatrix{ \lambda ^4 & \lambda ^3 \cr
            \lambda ^1 & \lambda ^0 }\;,
\label{d23new5}
\eea
from the $U(1)$ sector. The contributions from the full $U(1)\times Z_2$ 
symmetry give the Yukawa matrix,
\bea
&&{\bf D}\sim \pmatrix{ \lambda ^5 & \lambda ^3 \cr
            \lambda ^1 & \lambda ^1 }\;.
\label{d23new6}
\eea
The eigenvalues of this matrix are $\lambda ^0$ and $\lambda ^2$ and the 
mixing angle for the left-handed diagonalization matrix is of order
$\lambda ^2$. So one can interpret the [1,1] entry of order $\lambda ^5$
as being phenomenologically irrelevant to leading order in powers of 
$\lambda$ and the texture zero pattern is 
\bea
\pmatrix{0 & X \cr X & X}\;.
\eea
Note that the [2,2] entry is not phenomenologically irrelevant, and is 
still denoted by $X$. 

Having generated the Yukawa matrices ${\bf U}$ and ${\bf D}$
in Eqs.~(\ref{u23new3}) and (\ref{d23new6}), we can account for
phenomenological requirements (the full set of experimental data for fermion
masses and mixings is given in the next section),
\bea
&&{{m_c}\over {m_t}}\sim \lambda ^4\;, \qquad
{{m_s}\over {m_b}}\sim \lambda ^2\;,\qquad
|V_{cb}|\sim \lambda ^2\;.\label{2gendata} 
\eea
The mixing angles for ${\bf U}$ are $\sin \theta ^u_L\sim \lambda ^4$ and
$\sin \theta ^u_R\sim \lambda ^2$, while for ${\bf D}$ they are
$\sin \theta ^d_L\sim \lambda ^2$ and
$\sin \theta ^d_R\sim \lambda ^0$. The leading order contribution to 
$|V_{cb}|$ according to Eq.~(\ref{ckmdef}) is then given entirely by 
$\sin \theta ^d_L\sim \lambda ^2$ since $\sin \theta ^u_L$ is suppressed  
by a relative factor of $\lambda ^2$. The mass eigenvalue ratios in 
Eq.~(\ref{2gendata}) are properly accounted for. So phenomenologically 
viable Yukawa matrices can be found with texture zeros, and these zeros 
reduce the number of unknown order one coefficients that contribute to 
masses and mixing angles at the leading order in $\lambda$.

An important feature of the $U(1)\times Z_2$ flavor symmetry is that one can 
achieve different leading order contributions to the left-handed mixing angles
in $V_u^L$ and $V_d^L$, as shown in the above example. In a model with 
a $U(1)$ symmetry, these mixing angles are determined entirely by the 
charges $Q_L$ (and not $\overline{u}_R$ and $\overline{d}_R$). So the presence
of the $Z_2$ symmetry allows one to suppress the contribution to the CKM 
mixings from either the ${\bf U}$ or the ${\bf D}$ matrix.
  
It is not difficult to generalize the discussion to an arbitrary $Z_m$.
The exponents of $\lambda$ are given by Eq.~(\ref{u23}), and 
the conditions satisfied by the $Z_m$ charges that lead 
to texture suppressions are 
\bea
&&[q^Z_i+u^Z_i]+[q^Z_j+u^Z_j]-[q^Z_i+u^Z_j]-[q^Z_j+u^Z_i]=\pm m\;, 
\qquad m\ge 2\;, 
\label{cond}
\eea
where the case $+m$ results in a suppression of the diagonal entries of the 
$2\times 2$ matrix, and the case $-m$ results in a suppression of the 
off-diagonal entries. We remind the reader that the square brackets in 
Eq.~(\ref{cond}) indicate a modding by the integer $m$.

\subsection{Extending $Z_m$-induced suppressions to $3\times 3$ matrices}

In the last subsection examples of how to suppress entries in a $2\times 2$ 
Yukawa matrix were presented. One can extend this result to 
the three generation case by considering $2\times 2$ blocks. One has
three such blocks in the case of three generations: namely the [2-3], [1-3]
and the [1-2] blocks. One can build up a texture pattern for a $3\times 3$ 
matrix by placing zeros in the desired positions of these $2\times 2$ blocks.
As demonstrated in the last subsection, in each $2\times 2$ block, one can 
have either a texture zero in the off-diagonal or in a diagonal position, but 
not both at the same time.
As an example consider the matrix:
\begin{equation}    
\left (\begin{array}{ccc}
         0 & 0 & X \cr
         0 & X & X \cr
         X & X & X
\end{array} \right)\label{2te}
\end{equation}
All the zeros cannot be 
obtained by a assigning charges in the [1-2] block alone, since this would 
require the zeros to be in both the diagonal and off-diagonal positions.
However this texture pattern can be obtained by assigning the zero on the 
diagonal to the [1-3] block, while off-diagonal 
zeros can be assigned to the [1-2] block. As in the case of only 
two generations, one can obtain the texture pattern by considering only the
$Z_m$ component of the flavor symmetry. One can obtain
the required texture in Eq.~(\ref{2te}) when $m=3$ by the following assignment 
of $Z_3$ charges
\bea
&& Q_L: \quad q^Z_i=(2,0,1)\;, \qquad \overline{u}_R: \quad u^Z_i=(2,0,1)\;. 
\qquad i=1,2,3\;, \label{particular5}
\eea
The contribution to the $3\times 3$ matrix Yukawa matrix from this $Z_3$ 
charge assignment is
\bea
&&\pmatrix{\lambda^1 & \lambda^2 & \lambda^0\cr
                    \lambda^2 & \lambda^0 & \lambda^1 \cr
                    \lambda^0 & \lambda^1 & \lambda^2}\;,
\label{ex-texture}
\eea
The $U(1)$ 
contributions have not been included yet in Eq.~(\ref{ex-texture}). 
While the [2-3] 
block does not have a suppressing pattern, the suppression in the 
[1-3] block suppresses the 
diagonal element [1,1]. Finally,
the [1-2] block suppresses the off-diagonal 
[1,2] and [2,1] elements. 

Continuing with our example, if we assign the $U(1)$ flavor charges
\bea
&& Q_L: \quad q_i=(6,3,0)\;, \qquad \overline{u}_R: \quad u_i=(6,3,0)\;. 
\qquad i=1,2,3\;, \label{particular6}
\eea
to the quark fields, we obtain the following contribution to the 
up-type Yukawa matrix,
\bea
\pmatrix{\lambda ^{12} & \lambda ^9 & \lambda ^6 \cr
                 \lambda ^9    & \lambda ^6 & \lambda ^3 \cr
                 \lambda ^6 & \lambda ^3 & 1}\;.
\eea
Putting the contributions from both components of the $U(1)\times Z_3$ flavor
symmetry together gives the following up-type Yukawa matrix (after dropping an
overall factor of $\lambda ^2$ which is irrelevant as far as the hierarchy
is concerned), 
\bea
{\bf U}&&\sim \pmatrix{\lambda ^{11} & \lambda ^9 & \lambda ^4 \cr
                 \lambda ^9    & \lambda ^4 & \lambda ^2 \cr
                 \lambda ^4 & \lambda ^2 & 1}\;. \label{diagonal1}
\eea
One can always diagonalize matrices arises from Abelian flavor symmetries
of the type described here in stages\cite{hr}, by 
diagonalizing the [2-3] block, followed by the [1-3] block, and finally 
diagonalizing the [1-2] block. The diagonalization in the [2-3] block does
not produce any texture zero because $(\lambda ^2)(\lambda ^2)/(1)(\lambda ^4)$
as in Eq.~(\ref{rule}). Each order one coefficient in the [2-3] block
plays a role in determining the leading order diagonalization of that block.
However in the diagonalization of the [1-3] block,
one notices that [1,1] element ($\lambda ^{11}$) is suppressed by a factor 
of order $\lambda ^3$ compared to the product of the [1,3] and [3,1] elements.
So to leading order in an expansion in $\lambda $, the diagonalization 
of the matrix in Eq.~(\ref{diagonal1}) is the same as a matrix where the
$\lambda ^{11}$ element is replaced with zero (and we call such an entry a 
texture zero). So we have the following matrix whose diagonalization is 
equivalent to leading order to the original matrix
\bea
{\bf U}&&\sim \pmatrix{ 0  & \lambda ^9              & \lambda ^4 \cr
               \lambda ^9   & \lambda ^4 & \lambda ^2 \cr
                 \lambda ^4 & \lambda ^2 & 1}\;. \label{diagonal2}
\eea
Finally we must determine if any of the elements in the [1-2] block are
suppressed. Suppose the diagonalization has been performed in the [2-3] and 
[1-3] blocks. Then the matrix has the form 
\bea
&&\pmatrix{   \lambda^8       & \lambda ^9 & 0 \cr
                 \lambda^9   & \lambda ^4 & 0 \cr
                 0 & 0 & 1}\;.\label{diagonal2b}
\eea
The $\lambda ^8$ entry would be generated in the [1,1] element (and the 
$\lambda ^{11}$ element can be neglected in comparison, as described above). 
So a subsequent diagonalization of the [1-2] block indicates that  
texture zeros occur in the off-diagonal elements as
\bea
{\bf U}&&\sim \pmatrix{   0       & 0              & \lambda ^4 \cr
                 0   & \lambda ^4 & \lambda ^2 \cr
                 \lambda ^4 & \lambda ^2 & 1}\;.\label{diagonal3}
\eea
In other words, to leading order in $\lambda$ the diagonalization of the first
matrix in Eq.~(\ref{diagonal1}) is the same as the diagonalization of the 
matrix in Eq.~(\ref{diagonal3}).

By proceeding in this way, one can systematically construct all possible 
matrices with texture zeros in the desired positions. The task then is to 
combine a texture pattern for the up-type Yukawa matrix with another texture
pattern for the down-type Yukawa matrix, and check whether 
all the phenomenological requirements can be satisfied. We now turn 
to the experimental data for the quark and lepton masses and mixing angles.

\section{Phenomenological Requirements in the Quark Sector}

If one must satisfy the phenomenological constraints with positive 
flavor charges, then the 
Eq.~(\ref{eir}) is
the solution that results from a $U(1)$ flavor symmetry. Using a 
$U(1)\times Z_m$ flavor symmetry instead will change the exponents by adding 
$m$ in certain elements. The relevant equations for the CKM matrix elements
that are valid for this category of matrices are\cite{bk},
\bea
|V_{us}|&=&\left ({{d_{12}}\over {\tilde{d}_{22}}}
-{{d_{13}d_{32}}\over 
{\tilde{d}_{22}}}\right )
-\left ({{u_{12}}\over {\tilde{u}_{22}}}-{{u_{13}u_{32}}\over 
{\tilde{u}_{22}}}\right )\;, \label{Vusangles} \\
|V_{cb}|&=&d_{23}+d_{22}d_{32}^*-u_{23}\;, \label{Vcbangles} \\
|V_{ub}|&=&(d_{13}+d_{12}d_{32}^*-u_{13})-\left ({{u_{12}}\over 
{\tilde{u}_{22}}}-
{{u_{13}u_{32}}\over {\tilde{u}_{22}}}\right )(d_{23}+d_{22}d_{32}^*-u_{23})\;,
\label{Vubangles} \\
|V_{td}|&=&-(d_{13}+d_{12}d_{32}^*-u_{13})+\left ({{d_{12}}\over 
{\tilde{d}_{22}}}-
{{d_{13}d_{32}}\over {\tilde{d}_{22}}}\right )(d_{23}+d_{22}d_{32}^*-u_{23})\;,
\label{Vtdangles}
\eea
where $d_{ij}={\bf D}_{ij}/{\bf D}_{33}$ and 
$\tilde{d}_{22}=d_{22}-d_{23}d_{32}$
and $\tilde{u}_{22}=u_{22}-u_{23}u_{32}$.
It is understood that there will in general be relative phases between the 
terms on the right hand sides of Eqs.~(\ref{Vusangles})-(\ref{Vtdangles}), 
which are the correct forms to evaluate the leading orders 
for Yukawa matrices of the form considered in this paper. 

Taking the expansion parameter to be the 
Cabibbo angle, $\lambda=|V_{us}|$, 
then the experimental constraints\cite{pdg}
\bea
&&|V_{us}|=0.2196\pm 0.0023\;, \ \ |V_{cb}|=0.0395\pm 0.0017\;, \ \ 
\left |{{V_{ub}}\over {V_{cb}}}\right |=0.08\pm 0.02 \;,
\label{pdgdata}
\eea
on the CKM matrix can be identified in terms of powers of $\lambda$ by 
the following\footnote{There are renormalization scaling factors that relate
the experimental data at the electroweak scale, Eq.~(\ref{pdgdata}), to 
the relationships at the high scale\cite{bbo}.},
\bea
&&|V_{us}|\sim \lambda \;, \ \ |V_{cb}|\sim \lambda ^2 \;, \ \ 
|V_{ub}| \sim
\lambda ^3 -\lambda ^4\;, \ \ 
\left |{{V_{ub}}\over {V_{cb}}}\right |\sim \lambda -\lambda^2 \;.
\label{ckmelem}
\eea
We consider a model of Yukawa matrices to describe the experimental data 
satisfactorily if the leading order contribution to the CKM elements agrees 
with Eq.~(\ref{ckmelem}). For $|V_{ub}|$ and $|V_{ub}/V_{cb}|$ we accept
two values for the exponent of the leading contribution.
The constraint on $|V_{ub}/V_{cb}|$ can be expressed 
in a stronger way at 90\% confidence level as $0.25\lambda -0.5\lambda$.
One also has a constraint on 
the CKM elements from $B_d^0-\overline{B}_d^0$ mixing\cite{pdg},
\bea
&&|V_{tb}^*V_{td}|=0.0084\pm 0.0018\;, \label{pdgdata2}
\eea
which implies that 
\bea
&&|V_{td}|\sim \lambda ^3\;.\label{Vtd}
\label{ckmelem2}
\eea

The eigenvalues of the Yukawa matrices are constrained by the following
requirements from experimental observations
\bea
&&{{m_c}\over {m_t}}\sim \lambda ^4, \quad
  {{m_u}\over {m_c}}\sim \lambda ^4, \quad
  {{m_s}\over {m_b}}\sim \lambda ^2, \quad
  {{m_d}\over {m_s}}\sim \lambda ^2\;.
\label{masses}
\eea

These phenomenological requirements will be used in the 
next section to constrain the Yukawa matrix patterns that
can successfully reproduce the experimental data.

\section{Grand Unified Model}

In this section
we derive an assignment of charges in $U(1)\times Z_2$ that has the 
maximum number of texture suppressions (four) that is consistent with 
a $SU(5)$ grand unified symmetry. Since the flavor symmetry is required to 
commute with $SU(5)$, this means that there must be a common flavor charge 
assignment for all particles in each multiplet of $SU(5)$.
We restrict our attention to the case of a $Z_2$ symmetry, since (as described
earlier) it is the only possible $Z_m$ symmetry that can reproduce a 
hierarchy in neutrino masses of order $\lambda ^2$\cite{gns}.

Firstly, we have found that all the solutions from the 
$U(1)\times Z_2$ flavor symmetry that satisfy the quark sector 
phenomenology have the following property: the [3,2] entry and
the [3-3] entry of the down quark Yukawa matrix, ${\bf D}$,
are the same order of
magnitude. If the flavor symmetry is embedded in a grand unified model, the
charged lepton Yukawa matrix will be given by the transpose of ${\bf D}$.
Then the feature of Eq.~(\ref{eir}), that the right-handed mixing matrix that
diagonalizes the down-quark Yukawa matrix, {\bf D}, is of 
order one in the [2-3]
block, is retained. 
This has the important consequence that, if the lepton charges are 
related to the down quark charges by a grand unified theory, then the charged
lepton matrix will require a large mixing between in the [2-3] block to 
diagonalize it. This results in a large mixing between the second and third
generation of neutrinos, and can naturally explain how the atmospheric 
neutrino mixing can be large (order one) while the quark mixing between the
second and third generations, $|V_{cb}|$, can be small (order $\lambda ^2$).
This has been called the ``lopsided'' solution to the producing the required
atmospheric neutrino mixing in grand unified models\cite{barr}.
This occurs in all the models necessarily after applying 
the phenomenological requirements
$|V_{cb}|\sim \lambda ^2$ and $m_s/m_b \sim \lambda ^2$.

In models in which the $U(1)$ 
flavor symmetry is gauged and anomalous, one can imagine the anomaly is 
canceled via the Green-Schwarz mechanism\cite{gs}. A convenient 
way to ensure that the flavor charges are amenable to cancellation is to have
the flavor symmetry commute with the $SU(5)$ grand unified 
theory\footnote{The mixed Standard Model-$U(1)$ anomalies can be canceled 
entirely by the Green-Schwarz mechanism  if the $U(1)$ charges $X$ satisfy 
the relations $tr(XT_aT_b)\propto tr(T_aT_b)$ and $tr(X^2Y)=0$
where $T_a$ are the Standard Model generators. These relations are 
satisfied automatically if the $U(1)$ charges respect the $SU(5)$ symmetry.}.
In the traditional $SU(5)$ grand unified theory, 
the fields $Q_L$ and $\overline{u}_R$ are assigned to the ${\bf 10}$ 
representation, and the $\overline{d}_R$ is assigned to the ${\bf 5^*}$ 
representation. 
We have found a texture pattern for the up and down quark 
Yukawa matrices with four texture zeros
for which the flavor symmetry quantum number assignment commutes
with an $SU(5)$ grand unified gauge group.
This texture pattern yields 
\bea
\quad{\bf U}\sim \pmatrix{\lambda ^8 & 0 & 0\cr
                    0 & \lambda^4 & \lambda ^2\cr
                    0 & \lambda ^2 & 1} \quad
{\bf D}\sim \pmatrix{ 0 & 0 & \lambda ^3\cr
          0 & \lambda^2 & \lambda ^2\cr
                    \lambda & 1 & 1} &&
\quad\matrix{
Q_L:&(4,1)&(2,0)&(0,0)\cr
\overline{u}_R:&(4,1)&(2,0)&(0,0)\cr
\overline{d}_R:&(2,0)&(1,0)&(0,1)}\;.\label{ae}
\eea
This assignment has common $U(1)\times Z_2$
flavor symmetry quantum numbers for the $Q_L$ and $\overline{u}_R$ fields 
in the ${\bf 10}$, and a systematic search reveals that 
no other texture pattern with four or more texture
zeros satisfies this property.
Finding an assignment for which the flavor symmetry commutes with SU(5) allows
us to assign flavor charges to the rest of the $SU(5)$ multiplets, namely the
charged leptons and neutrinos. 

The texture pattern given by Eq.~(\ref{ae})
has the following feature: The CKM mixing
$|V_{cb}|$ arises from contributions of order $\lambda ^2$ from the 
diagonalizations of both the ${\bf U}$ and ${\bf D}$ Yukawa matrices. All
other CKM mixing angles ($|V_{us}|$, $|V_{ub}|$ and $|V_{td}|$)
arise solely from the ${\bf D}$ Yukawa matrix.

Given the quantum number assignment in Eq.~(\ref{ae}), we can extend the 
model to encompass the leptons. The field $\overline{e}_R$ fills out the 
${\bf 10}$ representation, and the left-handed lepton doublet, $L_L$, fills 
out the ${\bf 5^*}$ representation, so they should have the quantum number 
assignments
\bea
\matrix{i=&1&2&3\cr
\overline{e}_R:&(4,1)&(2,0)&(0,0)\cr
L_L:&(2,0)&(1,0)&(0,1)}\;,\label{aelepton}
\eea
These assignments dictated by the Eq.~(\ref{ae}) can be compared against 
constraints obtained from experiment for masses and mixings in the lepton 
sector.
The first phenomenological constraints we consider
involve the charged leptons. Using the $U(1)\times Z_2$ quantum numbers in
Eq.~(\ref{aelepton}), one immediately obtains the charged lepton Yukawa 
matrix (see the Appendix for formulae), 
\bea 
m_{\ell^\pm} \sim \pmatrix{\lambda ^7 & \lambda ^4 
                                             & \lambda ^2 \cr
                      \lambda ^6 & \lambda ^3 
                                             & \lambda  \cr
                      \lambda  ^4 & \lambda ^3 
                                             & \lambda }v_1\;.
\eea
As desired the [2,3] and [3,3] elements are the same order of magnitude.
This yields the mass ratios
\bea
{{m_\mu}\over {m_\tau}}\sim\lambda^2,
\ \ \ {{m_e}\over {m_\tau}}\sim\lambda^4\;, \label{chleptons}
\eea
which are consistent with the experimental constraints after including 
renormalization group scaling\cite{bbo}\footnote{The largest scaling effect
results from the additional running necessary to reach the muon and the 
electron mass scales so that one can relate the Yukawa couplings to the 
physical masses of the charged leptons. The scaling of the Yukawa coupling
ratios themselves is negligibly small.}.

Next consider the light neutrino mass matrix. There are two possibilities that
were discussed previously in Ref.~\cite{bk}. First the light neutrino mass 
matrix might not have suppressed entries arising 
from the $Z_2$ component of the 
flavor symmetry, in which case the light neutrino mass matrix is simply
given by Eq.~(\ref{lightnu}),
where $L_i$ in this case 
is simply the sum of the $U(1)$ and $Z_2$ quantum numbers of the 
relevant lepton doublet field, $L_L$.
For the charge assignments in Eq.~(\ref{aelepton}), this gives
\bea 
m_\nu \sim \pmatrix{\lambda ^4 & \lambda ^3 
                                             & \lambda ^3 \cr
                      \lambda ^3 & \lambda ^2 
                                             & \lambda ^2  \cr
                      \lambda  ^3 & \lambda ^2 
                                             & \lambda ^2 }
{{v_2^2}\over {\Lambda_L}}\;.
\eea

The remaining constraints on leptons involve the neutrino masses and mixings.
The most interesting aspect of the neutrino data is that the atmospheric 
neutrino mixing appears to be large, perhaps even maximal. As mentioned
earlier, it is difficult
to understand a hierarchical pattern for the neutrino masses, since large
mixing should result when the neutrino masses are of roughly the same order 
of magnitude. The Super-Kamiokande data\cite{superk} suggest that 
\bea
&&\Delta m^2_{23}\sim 2.2\times 10^{-3}~{\rm eV}^2\;, \quad
\sin ^2 2\theta_{23}^\nu \sim 1\;, \label{atmos}
\eea
where the subscripts indicate the generations of neutrinos involved in the 
mixing (we assume the mixing is between $\nu_\mu$ and $\nu_\tau$, and not
some sterile neutrino).

The solar neutrino flux can be explained by one of three distinct solutions.
Two of these involve matter-enhanced oscillation (MSW), while the third
involves vacuum oscillations (VO). The two MSW solutions are differentiated by
the size of the mixing angle, so one is usually called the small mixing angle
(SMA) solution, and the other is called the large mixing angle (LMA) solution.
The values required for the mixing parameters in each of these three cases
are shown in the table below.
\bea
\matrix{&\Delta m_{1x}^2\ [eV^2]&\sin^22\theta_{1x}\nonumber \\ \nonumber
{\rm MSW(SMA)}&5\times10^{-6}&6\times10^{-3}\\ \nonumber
{\rm MSW(LMA)}&2\times10^{-5}&0.8\\ \nonumber
{\rm VO}&8\times10^{-11}&0.8}
\eea

The MSW solutions can be obtained with a $Z_2$ horizontal 
symmetry\cite{gns,bk,b-lepto}. If the neutrino masses are arranged in a 
hierarchy, then the best fit to the data is 
\bea
{{\Delta m^2_{12}\over\Delta m^2_{23}}\sim\lambda^4,
\ \ \ \sin \theta_{23}^\nu \sim \lambda^0,}
\eea
and either 
\bea
\sin \theta_{12}^\nu \sim\lambda^2,
\eea
for the SMA solution, or
\bea
\sin \theta_{12}^\nu \sim\lambda^0,
\eea
for the LMA solution.
If no $Z_2$ symmetry is operative, one gets a light neutrino mass as in 
Eq.~(\ref{lightnu}), and if $L_2=L_3$, there is no natural explanation for 
the hierarchy in the masses of $m_{\nu_\mu}$ 
and $m_{\nu_\tau}$. As explained in Refs.~\cite{bk,b-lepto},  
this can be remedied by assigning the right-handed neutrino fields 
$\overline{\nu}_{Ri}$ 
(singlets of $SU(5)$) the following $Z_2$ charges: (0,0,1). The particular 
$U(1)$ assignment for the fields $\overline{\nu}_{Ri}$ 
does not affect the light neutrino 
mass matrix.
In this case, the [3,3] element of the $m_\nu$ matrix is enhanced by a 
factor $\lambda ^{-2}$, giving 
\bea 
m_\nu \sim \pmatrix{\lambda ^4 & \lambda ^3 
                                             & \lambda ^3 \cr
                      \lambda ^3 & \lambda ^2 
                                             & \lambda ^2  \cr
                      \lambda  ^3 & \lambda ^2 
                                             & 1 }{{v_2^2}\over {\Lambda_L}}\;,
\label{aelightnu}
\eea
for the charge assignment in Eq.~(\ref{aelepton}).

The neutrino mixing matrix is 
\bea 
\pmatrix{1& \lambda  & \lambda  \cr
                   \lambda & 1 & 1  \cr
                      \lambda   & 1 & 1 }\;.
\eea
The solar mixing angle is predicted to be of order $\lambda$,
falling in between the optimal value for the LMA solution ($\lambda ^0$) and 
the SMA solution ($\lambda ^2$). Equation~(\ref{eir}) yields a solar mixing
angle of order $\lambda ^3$, so the presence of the $Z_2$ symmetry has the 
effect in the neutrino sector of enhance the mixing of the first generation to
the second and third generations by a factor $\lambda ^{-2}$. Several unknown
order-one
coefficients combine to produce the matrix in Eq.~(\ref{aelightnu}), so it is 
not necessarily inconsistent with the MSW solutions.

In the models described here, one can achieve alignment of the quark mass
matrices and the squark mass-squared matrices by certain positioning of 
texture zeros in the quark Yukawa matrices. This alignment can solve the 
SUSY flavor problem by making it possible to simultaneously diagonalize
the quark mass matrices and the quark-squark-gluino coupling, thereby 
avoiding the dangerous flavor-changing couplings. In particular, 
in the models we are discussing here, one can achieve this alignment if
there are texture zeros in the down quark Yukawa matrix, ${\bf D}$, in 
the [1,2] and [2,1] elements, and in either the [1,3] or [3,2] elements, and
in either the [1,2] or [3,1] elements. This is easily seen to be the 
case after a quick inspection of Eqs.~(\ref{Vusangles})-(\ref{Vtdangles}):
in this case the Cabibbo angle, $|V_{us}|$, arises to leading order solely in
the up quark Yukawa matrix, ${\bf U}$. The texture patterns
that achieve this alignment occur when the down-quark Yukawa matrix has 
texture zeros in the positions given by the patterns
\bea
&&    \pmatrix{X & 0 & 0 \cr
                0 & 0 & X \cr
                0 & X & X}\label{B}\\ \nonumber \\
&& \pmatrix{0 & 0 & X \cr
                0 & X & 0 \cr
                X & 0 & X}
\label{C}
\eea
which have the off-diagonal elements in the [1-2] block
doubly suppressed. The off-diagonal suppression in the [1-3] block
in the case of pattern Eq.~(\ref{B}) or the [2-3] block in the case of 
the pattern Eq.~(\ref{C}) need to be doubly suppressed, which is impossible.
So one cannot achieve the quark-squark
alignment in the context of a $U(1)\times Z_2$ flavor symmetry. 
On the other hand, one can employ the idea of supersymmetry breaking through 
an anomalous flavor symmetry\cite{dp,bd2} to the grand unified model 
presented in this section. One can obtain reasonable suppression of the 
flavor-changing effects provided the first and second generation sparticles
are in the multi-TeV range\cite{mr,nw,zhang}.

\section{Conclusion}
We have shown that if the fermion mass matrices are dictated by an Abelian 
family symmetry, one can obtain a phenomenologically successful texture 
pattern by employing additional $Z_m$ horizontal symmetries. 
This four-texture zero model has a flavor symmetry that commutes
with an SU(5) grand unified theory with the usual assignment of particles
to the ${\bf 5^*}$ and ${\bf 10}$ representations. When the quantum numbers
are extended to the lepton sector, the charged
lepton mass ratios were correctly predicted and a large mixing angle 
naturally arises to explain 
the atmospheric neutrino data. A mixing angle of order $\lambda$ arises
to explain 
the solar neutrino oscillation data.

Discrete flavor symmetries can suppress entries in the Yukawa matrices and
offer the potential of a solution to the supersymmetric flavor problem.
A judicious choice of texture zeros can render the quark mass matrices and 
the squark mass-squared matrices simultaneously diagonalizable, thereby 
eliminating some strongly constrained flavor-changing couplings. However,
we find that this solution cannot be obtained in a model with a 
single $Z_2$ symmetry and satisfy all the other (masses and mixings)
phenomenological requirements. However the quantum number assignments can 
be compatible through suppression of flavor-changing effects when supersymmetry
breaking is mediated by the anomalous flavor symmetry.

\section*{Appendix}

In this appendix we review the formulae for Yukawa and mass matrices that 
result from Abelian horizontal symmetries with and without a discrete
component. Let the $U(1)$ quark charges be given by 
\bea
&&\begin{array}{c@{\quad}c@{\quad}c@{\quad}c@{\quad}c@{\quad}c
@{\quad}c@{\quad}c@{\quad}c}
Q_{L1} & Q_{L2} & Q_{L3} & \overline{u}_{R1} & \overline{u}_{R2} & 
\overline{u}_{R3} 
& \overline{d}_{R1} & \overline{d}_{R2} & \overline{d}_{R3} \\
           q_1 & q_2 & q_3 & u_1 & u_2 & u_3 & d_1 
                    & d_2 & d_3    \end{array}\;.\nonumber
\eea
Then the up and down quark Yukawa matrices, ${\bf U}$ and ${\bf D}$ are 
given by\footnote{We use the 
notation $a\sim b$ to indicate that 
$a$ and $b$ are the same order in $\lambda$.}
\bea
&&{\bf U}\sim \pmatrix{\lambda ^{q_1+u_1} & \lambda ^{q_1+u_2} 
                                             & \lambda ^{q_1+u_3} \cr
                      \lambda ^{q_2+u_1} & \lambda ^{q_2+u_2} 
                                             & \lambda ^{q_2+u_3} \cr
                      \lambda  ^{q_3+u_1} & \lambda ^{q_3+u_2} 
                                             & \lambda ^{q_3+u_3}}\;,
\qquad{\bf D}\sim \pmatrix{\lambda ^{q_1+d_1} & \lambda ^{q_1+d_2} 
                                             & \lambda ^{q_1+d_3} \cr
                      \lambda ^{q_2+d_1} & \lambda ^{q_2+d_2} 
                                             & \lambda ^{q_2+d_3} \cr
                      \lambda  ^{q_3+d_1} & \lambda ^{q_3+d_2} 
                                             & \lambda ^{q_3+d_3}}\;.
\eea 
It is understood that these matrices have unknown coefficients multiplying 
each element. The contributions to each element arise from a different 
operator in Eq.~(\ref{fnmech}), so they are in general independent of each 
other. Since these coefficients are not correlated there is no 
reason to expect the Yukawa matrices to have a zero eigenvalue. 

To compare the predictions of flavor symmetries to these phenomenological 
constraints, one has to relate the CKM elements to the entries in the 
Yukawa matrices. The Yukawa matrices ${\bf U}$ and ${\bf D}$ can be 
diagonalized by biunitary
transformations
\begin{eqnarray}
{\bf U^{diag}}&=&V_u^L{\bf U}V_u^{R\dagger} \;, \\
{\bf D^{diag}}&=&V_d^L{\bf D}V_d^{R\dagger} \;. \label{biunit}
\end{eqnarray}
The CKM matrix is then given by 
\begin{equation}
V\equiv V_u^LV_d^{L\dagger } \;.
\label{ckmdef}
\end{equation}
The left-handed
transformation matrices $V_u^L$ and $V_d^L$ can be defined in 
terms of three successive rotations in the (2,3), (1,3) and (1,2) sectors. 
These rotation angles of the transformation matrices can be expressed in 
terms of the elements of the Yukawa matrices as follows\cite{lns,hr}
\bea
s_{12}^u&=&{{u_{12}}\over {\tilde{u}_{22}}}+{{u_{11}u_{21}^*}\over 
{|\tilde{u}_{22}^*|^2}}-{{u_{13}(u_{32}+u_{23}^*u_{22})}\over {\tilde{u}_{22}}}
-{{u_{11}u_{31}^*(u_{23}^*+u_{32}u_{22}^*)}\over {|\tilde{u}_{22}^*|^2}}\;, \\
s_{13}^u&=&u_{13}+u_{11}u_{31}^*+u_{12}(u_{32}^*+u_{22}^*u_{23})+u_{11}u_{21}^*
(u_{23}+u_{22}u_{32}^*)\;, \\
s_{23}^u&=&u_{23}+u_{22}u_{32}^*\;,\label{angles}
\eea
where $u_{ij}$ is the $i,j$th component of the up quark Yukawa matrix, 
${\bf U}/({\bf U})_{33}$,
and $\tilde{u}_{22}=u_{22}u_{33}-u_{23}u_{32}$.
There are corresponding expressions for the $s_{ij}^d$ in terms of the 
components of the down quark Yukawa matrix, ${\bf D}$ (which are slightly 
more complicated due to the fact that the (2,3) sector mixing in $V_d^R$ 
might be of
order one).
Clearly contributions to the CKM matrix elements can come from a number
of terms. In this paper we are 
interested in determining only the leading
order contribution(s) to the CKM angles and the fermion masses.

Assume now that the lepton fields have charges under a $U(1)$ family symmetry
\bea
&&\begin{array}{c@{\quad}c@{\quad}c@{\quad}c@{\quad}c@{\quad}c
@{\quad}c@{\quad}c@{\quad}c}
\overline{e}_{R1} & \overline{e}_{R2} & \overline{e}_{R3} & 
\ell_{L1} & \ell_{L2} & \ell_{L3} 
& \overline{\nu}_{R1} & \overline{\nu}_{R2} & \overline{\nu}_{R3} \\
           E_1 & E_2 & E_3 & L_1 & L_2 & L_3 & {\cal N} _1 
                    & {\cal N} _2 & {\cal N} _3    \end{array}\;.
\label{u1qn}
\eea
All the flavor charges are
non-negative so holomorphic zeros do not play a role. The only suppressed
entries will arise because of a discrete component in the flavor symmetry
via a mechanism described below.

Given lepton doublet charges $L_i$ and right-handed 
neutrino charges ${\cal N}_i$
one has the following pattern for the 
charged lepton matrix 
\bea
m_{\ell^\pm} \sim \pmatrix{\lambda ^{L_1+E_1} & \lambda ^{L_1+E_2} 
                                             & \lambda ^{L_1+E_3} \cr
                      \lambda ^{L_2+E_1} & \lambda ^{L_2+E_2} 
                                             & \lambda ^{L_2+E_3} \cr
                      \lambda  ^{L_3+E_1} & \lambda ^{L_3+E_2} 
                                             & \lambda ^{L_3+E_3}}v_1\;,
\label{chlepton}
\eea
and for the neutrino Dirac mass matrix 
\bea
&&m_D \sim \pmatrix{\lambda ^{L_1+{\cal N}_1} & \lambda ^{L_1+{\cal N}_2} 
                                             & \lambda ^{L_1+{\cal N}_3} \cr
                      \lambda ^{L_2+{\cal N}_1} & \lambda ^{L_2+{\cal N}_2} 
                                             & \lambda ^{L_2+{\cal N}_3} \cr
                      \lambda  ^{L_3+{\cal N}_1} & \lambda ^{L_3+{\cal N}_2} 
                                             & \lambda ^{L_3+{\cal N}_3}}v_2\;,
\label{neudirac}
\eea
We have defined there the VEVs of the Higgs coupling to the down- and up-type 
quarks to be $v_1$ and $v_2$, and one usually defines $\tan \beta =v_2/v_1$.
To determine the neutrino mixing angles one rotates to a basis
where the charged lepton matrix is diagonal. This will give a contribution
to the mixing in the light neutrino species.
The relevant mixing contributing to atmospheric neutrino oscillations
comes from the right hand side of the charge lepton matrix,
$\lambda ^{L_2+E_3}/\lambda ^{L_3+E_3}$.

The Majarona mass
matrix is obtained from the charges of the right-handed neutrino flavor 
charges ${\cal N}_i$ and a heavy scale we lable as $\Lambda _L$,
\bea
&&M_N \sim \pmatrix{\lambda ^{2{\cal N}_1} & \lambda ^{{\cal N}_1+{\cal N}_2} 
                            & \lambda ^{{\cal N}_1+{\cal N}_3} \cr
                    \lambda ^{{\cal N}_1+{\cal N}_2} & \lambda ^{2{\cal N}_2} 
                              & \lambda ^{{\cal N}_2+{\cal N}_3} \cr
    \lambda  ^{{\cal N}_1+{\cal N}_3} & \lambda ^{{\cal N}_2+{\cal N}_3} 
                             & \lambda ^{2{\cal N}_3}}\Lambda _L\;.
\eea
Then one obtains the following form for the light neutrino mass matrix via
the see-saw formula 
\be
m_\nu=m_D { 1 \over M_N} m_D^T \label{eqn1}\;, \label{seesaw}
\ee
where $m_D$ is the neutrino Dirac mass matrix. Then\cite{gns,lr},
\bea
&&m_\nu \sim \pmatrix{\lambda ^{2L_1} & \lambda ^{L_1+L_2} 
                                             & \lambda ^{L_1+L_3} \cr
                      \lambda ^{L_1+L_2} & \lambda ^{2L_2} 
                                             & \lambda ^{L_2+L_3} \cr
                      \lambda  ^{L_1+L_3} & \lambda ^{L_2+L_3} 
                                             & \lambda ^{2L_3}}
{{v_2^2}\over {\Lambda_L}}\;. \label{lightnu}
\eea
If $L_2=L_3$ one can obtain ${\cal O}(1)$ mixing in the 2-3 
sector\cite{gn}.
On the other hand, one fails to get a mass hierarchy between the second
and third generation, since the two mass eigenvalues for the second and third 
generations are both of order $\lambda ^{2L_3}$. 

A  discrete Abelian family symmetry can be
employed to enhance or suppress masses and mixing angle relative to the 
predictions 
obtained when the family symmetry is the continuous $U(1)$ symmetry, and
this idea was pursued further in specific models\cite{tan,bk}.
The discrete $Z_m$ symmetry can result in the enhancement of 
entries in the light neutrino mass matrix\cite{gns}, and this enhancement
is compatible with the neutrino seesaw mechanism\cite{b-lepto,bk}. For example,
if the $U(1)$ quantum numbers in Eq.~(\ref{u1qn}) are replaced by 
$U(1)\times Z_2$ quantum numbers, $L_3\rightarrow (L_3-1,1)$ and 
${\cal N}_3 \rightarrow ({\cal N}_3-1,1)$ so that the 
charges for the lepton 
fields are
\bea
&&\begin{array}{c@{\quad}c@{\quad}c@{\quad}c@{\quad}c
@{\quad}c@{\quad}c@{\quad}c@{\quad}c}
\overline{e}_{R1} & \overline{e}_{R2} & \overline{e}_{R3} 
& \ell_{L1} & \ell_{L2} & \ell_{L3} 
& \overline{\nu}_{R1} & \overline{\nu}_{R2} & \overline{\nu}_{R3} \\
           (E_1,0) & (E_2,0) & (E_3,0) & (L_1,0) & (L_2,0) & (L_3-1,1) 
& ({\cal N}_1,0) & ({\cal N}_2,0) & ({\cal N}_3-1,1)    \end{array}\;.
\eea
then one finds that 
\bea
&&M_N \sim \pmatrix{\lambda ^{2{\cal N}_1} & \lambda ^{{\cal N}_1+{\cal N}_2} 
                                & \lambda ^{{\cal N}_1+{\cal N}_3} \cr
                   \lambda ^{{\cal N}_1+{\cal N}_2} & \lambda ^{2{\cal N}_2} 
                                    & \lambda ^{{\cal N}_2+{\cal N}_3} \cr
          \lambda  ^{{\cal N}_1+{\cal N}_3} & \lambda ^{{\cal N}_2+{\cal N}_3} 
                                    & \lambda ^{2{\cal N}_3-2}}\Lambda _L\;,
\eea
so that 
\bea
&&(M_N)^{-1} \sim \pmatrix{\lambda ^{-2{\cal N}_1} & 
                  \lambda ^{-{\cal N}_1-{\cal N}_2} 
                              & \lambda ^{-{\cal N}_1-{\cal N}_3+2} \cr
              \lambda ^{-{\cal N}_1-{\cal N}_2} & \lambda ^{-2{\cal N}_2} 
                                   & \lambda ^{-{\cal N}_2-{\cal N}_3+2} \cr
   \lambda  ^{-{\cal N}_1-{\cal N}_3+2} & \lambda ^{-{\cal N}_2-{\cal N}_3+2} 
                               & \lambda ^{-2{\cal N}_3+2}}\Lambda _L^{-1}\;.
\eea
So the effect of the discrete symmetry in our case is to enhance the 3-3 entry 
of the $M_N$ matrix, and thereby alter the results for the third row and the 
third column on the inverse matrix, $(M_N)^{-1}$.
The 3-3 component
of the neutrino Dirac mass matrix is also enhanced by the discrete 
symmetry, so that Eq.~(\ref{neudirac}) is modified to be
\bea
&&m_D \sim \pmatrix{\lambda ^{L_1+{\cal N}_1} & \lambda ^{L_1+{\cal N}_2} 
                                             & \lambda ^{L_1+{\cal N}_3} \cr
                      \lambda ^{L_2+{\cal N}_1} & \lambda ^{L_2+{\cal N}_2} 
                                             & \lambda ^{L_2+{\cal N}_3} \cr
                      \lambda  ^{L_3+{\cal N}_1} & \lambda ^{L_3+{\cal N}_2} 
                                    & \lambda ^{L_3+{\cal N}_3-2}}v_2\;,
\eea
The light neutrino mass matrix in Eq.~(\ref{lightnu}) is modified so that 
only the 3-3 entry is enhanced,
\bea
&&m_\nu \sim \pmatrix{\lambda ^{2L_1} & \lambda ^{L_1+L_2} 
                                             & \lambda ^{L_1+L_3} \cr
                      \lambda ^{L_1+L_2} & \lambda ^{2L_2} 
                                             & \lambda ^{L_2+L_3} \cr
                      \lambda  ^{L_1+L_3} & \lambda ^{L_2+L_3} 
                                             & \lambda ^{2L_3-2}}
{{v_2^2}\over {\Lambda_L}}\;. \label{lightnuZ2}
\eea
The charged lepton mass matrix, Eq.~(\ref{chlepton}), and hence a large mixing
angle is needed to diagonalize the [2-3] block. So the large mixing observed in
the atmospheric neutrino experiments is accounted for, while the hierarchy
of order $\lambda ^2$ in the second and third generation neutrino masses is 
obtained. 
Generalizing to a discrete symmetry $Z_m$ rather than $Z_2$ 
can preserve the large neutrino
mixing while enhance the heaviest neutrino mass by a factor $\lambda ^{-m}$.
\section*{Acknowledgments}

This work was supported in part by the U.S.
Department of Energy
under Grant No. 
No.~DE-FG02-91ER40661.


\begin{thebibliography}{99}

\bibitem{superk} Y.~Fukuda, et al., The Super-Kamionkande Collaboration,
Phys.\ Rev.\ Lett.\ {\bf 81}, 1562 (1998).

\bibitem{gns} Y.~Grossman, Y.~Nir and Y.~Shadmi, J. High Energy Phys. {\bf 10},
007 (1998); Y.~Nir and Y.~Shadmi, {\it ibid.} 9905, 023 (1999).

\bibitem{gn} Y.~Grossman and Y.~Nir, Nucl.\ Phys.\ {\bf B448}, 30 (1995).

\bibitem{lns} M.~Leurer, Y.~Nir and N.~Seiberg, Nucl.\ Phys.\ {\bf B398}, 319 
(1993); Nucl.\ Phys.\ {\bf B420}, 468 (1994).

\bibitem{tan} M.~Tanimoto, Phys.\ Lett.\ {\bf B456}, 220 (1999).

\bibitem{bk} M.~S.~Berger and K.~Siyeon, Phys.\ Rev.\ {\bf D62} 
033004 (2000).

\bibitem{flavor} 
Z.~Berezhiani and A.~Rossi, Nucl.\ Phys.\ {\bf B594}, 113 (2001); 
Q.~Shafi and Z.~Tavartkiladze, 
Phys.\ Lett.\ {\bf B487}, 145 (2000); A.~Aranda, C.~D.~Carone and R. F.~Lebed, 
Phys. Lett.\ {\bf B474}, 170. 

\bibitem{hm} L.~J.~Hall and H.~Murayama, Phys.\ Rev.\ Lett.\ {\bf 77}, 
(1996) 1699; C.~D.~Carone, L.~J.~Hall and H.~Murayama, Phys.\ Rev.\ 
{\bf D53}, 6282 (1996).

\bibitem{fk} P.~H.~Frampton and O.~C.~Kong, Phys.\ Rev.\ Lett.\ {\bf 77}, 
(1996) 1699.

\bibitem{b-lepto} M.~S.~Berger, Phys.\ Rev.\ {\bf D62}, 013007 (2000).

\bibitem{ns} Y.~Nir and N.~Seiberg, Phys.\ Lett. {\bf B309}, 337 (1993).

\bibitem{fn} C.~D.~Froggatt and H.~B.~Nielsen, Nucl.\ Phys.\ {\bf B147}, 277 
(1979).

\bibitem{Ibanez} L.E.~Ibanez, Phys.\ Lett.\ {\bf B303}, 55 (1993).

\bibitem{ir} L.~Ibanez and G.~G.~Ross, Phys.\ Lett.\ {\bf B332}, 100 (1994).

\bibitem{blr} P.~Binetruy and P.~Ramond, Phys.\ Lett.\ {\bf B350}, 49 (1995);
P.~Binetruy, S.~Lavignac and P.~Ramond, Nucl.\ Phys.\ {\bf B477}, 353 (1996).

\bibitem{nir} Y.~Nir, Phys. Lett. {\bf B354}, 107 (1995).

\bibitem{dllrs} H.~Dreiner, G.~K.~Leontaris, S.~Lola, G.~G.~Ross and 
C.~Scheich, Nucl.\ Phys.\ {\bf B436}, 461 (1995).

\bibitem{bd} P.~Binetruy and E.~Dudas, Nucl.\ Phys.\  {\bf B442}, 21 (1995);
E.~Dudas, S.~Pokorski and C.~A.~Savoy, Phys.\ Lett.\  {\bf B356}, 45 (1995);
E.~J.~Chun and A.~Lukas, Phys.\ Lett.\  {\bf B387}, 99 (1996);
N.~Irges, S.~Lavignac and P.~Ramond, Phys.\ Rev.\  {\bf D58}, 035003 (1998).

\bibitem{mr} R.~N.~Mohapatra and A.~Riotto, Phys.Rev.\ {\bf D55}, 1138 (1997);
Phys.Rev.\ {\bf D55}, 4262 (1997).

\bibitem{nw} A.~E.~Nelson and D.~Wright, Phys.\ Rev.\ {\bf D56}, 1598 (1997).

\bibitem{zhang} R.-J.~Zhang, Phys.\ Lett.\ {\bf B402}, 101 (1997).

\bibitem{lr2} G.K.~Leontaris and J.~Rizos, Nucl.\ Phys.\ {\bf B567}, 32 
(2000). 

\bibitem{af} G.~Altarelli and F.~Feruglio, Phys.\ Lett.\ {\bf B451} 388 
(1999).

\bibitem{gs} M.~Green and J.~Schwarz, Phys.\ Lett.\ {\bf 149B}, 117 (1984).

\bibitem{lr} S.~Lola and G.~G.~Ross, Nucl.\ Phys.\ {\bf B553}, 81 (1999).

\bibitem{eir} J.~K.~Elwood, N.~Irges and P.~Ramond, Phys.\ Rev.\ Lett.\
{\bf 81}, 5064 (1998).

\bibitem{irges} N.~Irges, Phys.\ Rev.\ {\bf D59}, 115008 (1999).

\bibitem{hr} L.~J.~Hall and A.~Rasin, Phys.\ Lett.\ {\bf B315}, 164 (1993).

\bibitem{pdg} Review of Particle Physics, Euro.\ Phys.\ Jour.\ {\bf C3}, 1 
(1998).

\bibitem{bbo} V.~Barger, M.~S.~Berger and P.~Ohmann, Phys.\ Rev.\ {\bf D47},
1093 (1993); Phys.\ Rev.\ {\bf D47}, 2038 (1993).

\bibitem{barr} S.~Barr, hep-ph/0003101.

\bibitem{dp} G.~Dvali and A.~Pomarol, Phys.\ Rev.\ Lett.\ {\bf 77}, 3728 
(1996).

\bibitem{bd2} P.~Bin\'{e}truy and E.~Dudas, Phys. Lett.\ {\bf B389}, 503 
(1996).

\end{thebibliography}
\end{document}